\documentclass[superscriptaddress,twocolumn,showpacs,a4paper,
amssymb,amsmath,longbibliography,nobibnotes,aps,prd,
showkeys,
nofootinbib,notitlepage]{revtex4-1}
\usepackage{verbatim}
\usepackage[T1]{fontenc}
\usepackage[utf8]{inputenc}
\usepackage[american]{babel}
\usepackage{epsfig}
\usepackage{booktabs}
\usepackage{multirow}
\usepackage{dcolumn}
\usepackage{amsmath}
\usepackage{mathtools}
\usepackage{amsfonts}
\usepackage{amssymb}
\usepackage{ulem}
\usepackage{epstopdf}
\usepackage{bm}
\usepackage{siunitx}
\usepackage{braket}
\usepackage{enumitem}
\usepackage{soul}
\usepackage[table]{xcolor}
\usepackage{color}
\usepackage{transparent}
\usepackage{pifont}
\usepackage{enumitem}

\definecolor{navyblue}{rgb}{0.0, 0.0, 0.5}
\definecolor{royalblue}{rgb}{0.25, 0.41, 0.88}
\definecolor{cadmiumgreen}{rgb}{0.0, 0.42, 0.24}
\definecolor{blue-violet}{rgb}{0.54, 0.17, 0.89}
\definecolor{darkviolet}{rgb}{0.58, 0.0, 0.83}
\definecolor{orange(colorwheel)}{rgb}{1.0, 0.5, 0.0}

\usepackage{hyperref}
\hypersetup{
    colorlinks=true, 
    linkcolor=royalblue, 
    citecolor=blue}

\usepackage{booktabs}
\usepackage{multirow}
\usepackage{dcolumn}
\usepackage{colortbl}

\begin{document}

\title{An overview of what current data can (and cannot yet) say about evolving dark energy}

\author{William Giar\`e}
\email{w.giare@sheffield.ac.uk}
\affiliation{School of Mathematical and Physical Sciences, University of Sheffield, Hounsfield Road, Sheffield S3 7RH, United Kingdom}

\author{Tariq Mahassen}
\email{tmahassen1@sheffield.ac.uk}
\affiliation{School of Mathematical and Physical Sciences, University of Sheffield, Hounsfield Road, Sheffield S3 7RH, United Kingdom}

\author{Eleonora Di Valentino}
\email{e.divalentino@sheffield.ac.uk}
\affiliation{School of Mathematical and Physical Sciences, University of Sheffield, Hounsfield Road, Sheffield S3 7RH, United Kingdom}

\author{Supriya Pan}
\email{supriya.maths@presiuniv.ac.in}
\affiliation{Department of Mathematics, Presidency University, 86/1 College Street,  Kolkata 700073, India}
\affiliation{Institute of Systems Science, Durban University of Technology, PO Box 1334, Durban 4000, Republic of South Africa}

\date{\today}

\begin{abstract}

\noindent Recent measurements of Baryon Acoustic Oscillations (BAO) and distance moduli from Type Ia supernovae suggest a preference for Dynamical Dark Energy (DDE) scenarios characterized by a time-varying equation of state (EoS). This focused review assesses its robustness across independent measurements and surveys. Using the Chevallier-Polarski-Linder (CPL) parametrization to describe the evolution of the DE EoS, we analyze over 35 dataset combinations, incorporating Planck Cosmic Microwave Background (CMB) anisotropies, three independent Type Ia supernova (SN) catalogs (PantheonPlus, Union3, DESY5), BAO measurements from DESI and SDSS, and expansion rate measurements $H(z)$ inferred from the relative ages of massive, passively evolving galaxies at early cosmic times known as Cosmic Chronometers (CC). This review has two main objectives: first, to evaluate the statistical significance of the DDE preference across different dataset combinations, which incorporate varying sources of information. Specifically, we consider cases where only low-redshift probes are used in different combinations, others where individual low-redshift probes are analyzed together with CMB data, and finally, scenarios where high- and low-redshift probes are included in all possible independent combinations. Second, we provide a reader-friendly synthesis of what the latest cosmological and astrophysical probes can (and cannot yet) reveal about DDE. Overall, our findings highlight that combinations that \textit{simultaneously} include PantheonPlus SN and SDSS BAO significantly weaken the preference for DDE. However, intriguing hints supporting DDE emerge in combinations that do not include DESI-BAO measurements: SDSS-BAO combined with SN from Union3 and DESY5 (with and without CMB) support the preference for DDE.
\end{abstract}

\maketitle

\section{Introduction}
\label{sec:Introduction}

As first established in 1998 through observations of distant Type Ia Supernovae~\cite{SupernovaSearchTeam:1998fmf,SupernovaCosmologyProject:1998vns} and subsequently corroborated by numerous other probes~\cite{Sherwin:2011gv,Moresco:2016mzx,Rubin:2016iqe,Planck:2018vyg,Nadathur:2020kvq,Rose:2020shp,eBOSS:2020yzd}, our Universe is experiencing a phase of accelerated expansion. This groundbreaking discovery has significantly impacted our understanding of the Universe's ultimate fate, with profound implications for fundamental physics.

Explaining cosmic acceleration within the standard model of fundamental physics is highly challenging, as all known forces and components are expected to decelerate the expansion of the Universe. This enigma has inspired extensive theoretical exploration since the early 2000s, with cosmologists and particle physicists proposing diverse models and hypotheses. These range from vacuum energy density and new dynamical (pseudo-scalar) fields and interacting fluids to modified gravity theories extending General Relativity (GR)~\cite{Peebles:2002gy,Nojiri:2006ri,Copeland:2006wr,Durrer:2007re,Padmanabhan:2007xy,Bengochea:2008gz,Frieman:2008sn,Sotiriou:2008rp,DeFelice:2010aj,Clifton:2011jh,Li:2011sd,Bamba:2012cp,Koyama:2015vza,Berti:2015itd,Joyce:2016vqv,Wang:2016lxa,Wang:2016och,Nojiri:2017ncd,Bahamonde:2017ize,Bahamonde:2021gfp} (also see~\cite{Amendola:1999er,Huterer:2000mj,Boyle:2001du,Gorini:2002kf,Bagla:2002yn,Carroll:2003st,Carroll:2003wy,Nojiri:2003ft,Sahni:2002dx,Shapiro:2003ui,Espana-Bonet:2003qjh,Guo:2004fq,Simon:2004tf,Li:2004rb,Alam:2004jy,Nojiri:2004bi,Copeland:2004hq,Cai:2004dk,Wei:2005nw,Nojiri:2005vv,Capozziello:2005pa,Scherrer:2005je,Barrow:2006hia,Nojiri:2006gh,Amendola:2006we,Fay:2007uy,Tsujikawa:2007xu,Wei:2007ty,DeFelice:2008wz,Zhou:2009cy,Brax:2010gi,Harko:2011kv,Geng:2011aj,Ziaeepour:2011bq,Velten:2012uv,Bloomfield:2012ff,Lima:2012cm,Kamenshchik:2012pw,Perico:2013mna,Foffa:2013vma,Yang:2014hea,Linder:2015rcz,Pan:2012ki,Gomez-Valent:2014rxa,Shahalam:2015sja,Sola:2015wwa,Ballardini:2016cvy,Sola:2016jky,Brevik:2017msy,Rezaei:2017yyj,DiValentino:2017rcr,Saridakis:2017rdo,Yang:2017ccc,VanDeBruck:2017mua,SolaPeracaula:2017esw,Sola:2017znb,SolaPeracaula:2016qlq,Langlois:2017dyl,SolaPeracaula:2018wwm,Rezaei:2019xwo,Li:2019san,Anagnostopoulos:2018jdq,BeltranJimenez:2019tme,LinaresCedeno:2019bgo,Yang:2018xah,Pan:2019gop,Benetti:2019lxu,Akarsu:2019hmw,DiValentino:2019ffd,Yang:2020zuk,Granda:2020ikv,Cai:2021wgv,SolaPeracaula:2021gxi,Motta:2021hvl,SolaPeracaula:2022hpd,Ferrari:2023qnh,Yang:2023qqz,Halder:2024uao,SolaPeracaula:2023swx,Giare:2024smz,Gomez-Valent:2024tdb,Muralidharan:2024hsc,Giare:2024ytc,BarrosoVarela:2024htf,BarrosoVarela:2024ozs,Ghedini:2024mdu,Akarsu:2024nas,Koussour:2024jqp,Odintsov:2024woi,Halder:2024aan,Ferrari:2025egk,Halder:2025eze}).

Despite these efforts, the physical nature of the late-time accelerated expansion remains largely unknown. 
In the standard $\Lambda$CDM model, this phase of acceleration is attributed to a positive cosmological constant term ($\Lambda$) in Einstein's field equations. Nevertheless, unresolved issues with the cosmological constant~\cite{Weinberg:1988cp,Zlatev:1998tr} have necessitated consideration of alternative frameworks, including a dynamical fluid akin to $\Lambda$, known as dark energy (DE). The simplest DE model assumes a constant equation of state (EoS), $w_0$, which has been constrained using various observations, including the CMB from Planck and complementary experiments such as Atacama Cosmology Telescope (ACT) and South Pole Telescope (SPT)~\cite{Planck:2018vyg,ACT:2020gnv,SPT-3G:2021eoc,SPT-3G:2022hvq,Giare:2023xoc,Escamilla:2023oce}.\footnote{It is worth noting that Planck data show a mild preference for a phantom EoS ($w_0 < -1$)~\cite{Planck:2018vyg,Yang:2021flj,Escamilla:2023oce}, which is not confirmed by ACT or SPT~\cite{ACT:2020gnv,SPT-3G:2021eoc,SPT-3G:2022hvq,Giare:2023xoc}.}  However, there is no compelling reason (other than simplicity) to limit this model to a constant EoS.
Relaxing the assumption of a constant EoS leads naturally to considering a time-dependent EoS, $w(a)$, varying with the Universe's expansion~\cite{Cooray:1999da,Efstathiou:1999tm,Chevallier:2000qy,Linder:2002et,Wetterich:2004pv,Feng:2004ff,Hannestad:2004cb,Xia:2004rw,Gong:2005de,Jassal:2005qc,Nesseris:2005ur,Liu:2008vy,Barboza:2008rh,Barboza:2009ks,Ma:2011nc,Sendra:2011pt,Feng:2011zzo,Barboza:2011gd,DeFelice:2012vd,Feng:2012gf,Wei:2013jya,Magana:2014voa,Akarsu:2015yea,Pan:2016jli,DiValentino:2016hlg,Nunes:2016plz,Nunes:2016drj,Magana:2017usz,Yang:2017alx,Pan:2017zoh,Panotopoulos:2018sso,Yang:2018qmz,Jaime:2018ftn,Das:2017gjj,Yang:2018prh,Li:2019yem,Yang:2019jwn,Pan:2019hac,Tamayo:2019gqj,Pan:2019brc,DiValentino:2020naf,Rezaei:2020mrj,Perkovic:2020mph,Banihashemi:2020wtb,Jaber-Bravo:2019nrk,Benaoum:2020qsi,Yang:2021eud,Jaber:2021hho,Alestas:2021luu,Yang:2022klj,Escudero:2022rbq,Castillo-Santos:2022yoi,Yang:2022kho,Dahmani:2023bsb,Escamilla:2023oce,Rezaei:2023xkj,Adil:2023exv,LozanoTorres:2024tnt,Singh:2023ryd,Rezaei:2024vtg,Reyhani:2024cnr,Najafi:2024qzm,Giare:2024gpk,Giare:2024ocw}.

Recently, BAO measurements from the Dark Energy Spectroscopic Instrument (DESI)~\cite{DESI:2024uvr,DESI:2024kob}, in combination with Planck 2018 CMB~\cite{Planck:2018vyg} and supernova type Ia datasets, have suggested evidence for dynamical DE transitioning from a past phantom-like to a present quintessence-like behavior. This analysis, based on the Chevallier-Polarski-Linder (CPL) parametrization, $w(a) = w_0 + w_a(1-a)$~\cite{Chevallier:2000qy,Linder:2002et} -- where $w_0$ corresponds to the present-day DE EoS and $w_a$ describes the dynamical evolution of DE -- indicates $w_0 > -1$ and $w_a < 0$ at a significance of $2.5\sigma$ to $3.9\sigma$, depending on the specific supernova type Ia dataset used~\cite{DESI:2024mwx}. 
Given the importance of the result, its robustness has been extensively tested in the literature~\cite{Cortes:2024lgw,Shlivko:2024llw,Luongo:2024fww,Yin:2024hba,Gialamas:2024lyw,Dinda:2024kjf}, including different combinations of datasets~\cite{Wang:2024dka,Chan-GyungPark:2024spk,Giare:2024ocw,Reeves:2025axp} or different parametrizations for Dynamical DE including various EoS of DE or different DE models~\cite{Tada:2024znt,Carloni:2024zpl,Chan-GyungPark:2024mlx,DESI:2024kob,Bhattacharya:2024hep,Ramadan:2024kmn,Notari:2024rti,Orchard:2024bve,Hernandez-Almada:2024ost,Pourojaghi:2024tmw,Giare:2024gpk,Reboucas:2024smm,Giare:2024ocw,Chan-GyungPark:2024brx,Menci:2024hop,Li:2024qus,Li:2024hrv,Notari:2024zmi,Gao:2024ily,Fikri:2024klc,Jiang:2024xnu,Zheng:2024qzi,Gomez-Valent:2024ejh,RoyChoudhury:2024wri,Lewis:2024cqj,Wolf:2024eph,Wolf:2024stt,Wolf:2025jlc,Yang:2025kgc,Sakr:2025daj,Shajib:2025tpd}. 

Undoubtedly, the hints of a dynamical evolution of dark energy reported by DESI represent one of the most intriguing indications of new physics beyond $\Lambda$CDM, as they challenge the cosmological constant interpretation of DE. The implications extend from fundamental physics to the ongoing tensions between Planck results within the $\Lambda$CDM paradigm~\cite{Planck:2018vyg} and local measurements, such as SH0ES~\cite{Riess:2021jrx}, potentially marking the beginning of a new era in cosmology.

Given the significance of this result, a systematic analysis is essential before drawing firm conclusions. Motivated by this crucial issue, we investigate whether different independent datasets, considered in various combinations, consistently support a preference for dynamical DE. In this focused review, following the approach adopted by DESI, we assume a CPL parametrization to describe the evolution of $w(a)$ and explore the full range of available measurements from diverse astronomical probes in constraining the DE EoS. We systematically analyze combinations of surveys, including Planck Cosmic Microwave Background anisotropies, three independent Type Ia supernova catalogs (PantheonPlus, Union3, DESY5), BAO measurements from DESI and SDSS, and Cosmic Chronometers. 

We have here two main objectives:

\begin{itemize}
\item We evaluate the constraining power of each individual probe to determine whether, and to what extent, they support the preference for DDE both independently and in combination with other high- and low-redshift surveys. Specifically, we compute the statistical significance of the DDE preference across different dataset combinations, incorporating varying sources of information. We first analyze cases where only low-redshift probes are used in different combinations, then examine scenarios where individual low-redshift probes are combined with CMB data, and finally, explore configurations where high- and low-redshift probes are included in all possible independent combinations.

\item We provide the community with a clear, accessible, and reader-friendly synthesis of what the latest cosmological and astrophysical probes can (and cannot yet) reveal about DDE. This include assessing whether and to what extent the preference for DDE should be regarded as a robust feature emerging from multiple independent datasets or whether it primarily arises from specific dataset combinations.
\end{itemize}

This review is organized as follows. In Sec.~\ref{sec:Model}, we present the basic framework of a generalized DE model and introduce the main model under study. Sec.~\ref{sec:data+Methodology} details the observational datasets and analysis methodology. Results are discussed in Sec.~\ref{sec:Results}, and conclusions are provided in Sec.~\ref{sec:Conclusion}. 

\section{Dynamical Dark energy}
\label{sec:Model}

We consider that our universe is homogeneous and isotropic in the large scale. This geometrical set-up of the universe is well described by the spatially flat Friedmann-Lema\^{i}tre-Robertson-Walker (FLRW) line element characterized by $ds^2 = -dt^2 + a^2 (t) (dx^2 +dy^2 +dz^2)$, where $a(t)$ ($t$ denotes the cosmic time) is the expansion scale factor of the universe and $(x, y, z)$ are the spatial coordinates. Next, we assume that the gravitational sector of the universe is described by GR and the matter sector inside it is minimally coupled to gravity, i.e. GR and there is no interaction between any two fluids. With these prescriptions, using the Einstein's gravitational equations, one can determine the 
the evolution of the universe. 

Assuming that the matter sector of the universe comprises of radiation, pressure-less matter (baryons and cold dark matter), neutrinos and a DE fluid, the 
Hubble's equation can be written as:

\begin{eqnarray}
&& H^2 (a) = \frac{\kappa^2}{3}( \rho_{\rm r}+\rho_{\rm m}+\rho_{\nu}+\rho_{\rm DE}),\label{eqn-Hubble}
\end{eqnarray}
where $\rho_{\rm r}$, $\rho_{\rm m}$, $\rho_{\nu}$, $\rho_{\rm DE}$ are respectively the energy density of radiation, matter, neutrinos,\footnote{We note that here we fix the sum of neutrino mass to $\sum m_\nu=0.06$ eV, while the number of neutrino species is fixed to $N_{\textrm{eff}}=3.044$ in agreement with
the prediction from the standard model of particle physics. } and DE. Now, since the individual fluid obeys its own conservation equation $\dot{\rho}_i + 3 H (1+w_i) \rho_i =0$, in which $w_i = p_i/\rho_i$ ($\rho_i$  and $p_i$ are denoting the energy density and pressure of the $i$-th fluid, respectively) is the barotropic equation of state of the $i$-th fluid. Therefore,  solving the conservation equation for radiation ($w_{\rm r} = 1/3$), pressure-less matter ($w_{\rm m} =0$), one would respectively obtain, $\rho_{\rm r} = \rho_\textrm{r,0}(a/a_0)^{-4}$, 
$\rho_{\rm m} = \rho_\textrm{m,0}(a/a_0)^{-3}$, where $a_0$ refers to the present day value of the scale factor;\footnote{Without any loss of generality, one can set $a_0 =1$ and we have also set the same in this article.} $\rho_\textrm{r,0}$,  $\rho_\textrm{m,0}$ are respectively the present day value of $\rho_{\rm r}$ and $\rho_{\rm m}$. On the DE side, using its conservation equation, the evolution of its energy density can be expressed as 
\begin{eqnarray}\label{rho-DE}
  \rho_{\rm DE} =   \rho_{\rm DE,0} a^{-3} \exp \left(-3 \int_{1}^{a} \frac{w(a')}{a'} \, da' \right),
\end{eqnarray}
where $w(a)$ is the barotropic equation of state of DE which could be either constant or time-varying. For $w =-1$ which corresponds to the cosmological constant, using the above equation one gets back $\rho_{\rm DE} =   \rho_{\rm DE,0}$, i.e. a constant component that represents the present-day value of the DE density. 

Depending on the nature of the EoS of DE, one can solve the evolution equation for the DE density (\ref{rho-DE}) either analytically or numerically, and then using the Hubble equation (\ref{eqn-Hubble}), one can solve for the scale factor of the universe, that means the expansion of the universe can be determined at the background level. However, the background evolution alone does not reveal the full picture
A complete description of a cosmological model also requires an understanding of its evolution at the perturbation level. A time varying DE EoS can influence the perturbative evolution of the universe, directly by affecting the formation of structure of the universe. It is therefore essential to understand how a time varying DE EoS influences the universe's structure formation. This however needs a specific functional form for $w_{\rm DE}$.

In order to understand the overall picture at the level of perturbations, we consider the following line element written 
in the synchronous gauge~\cite{Ma:1995ey}
\begin{equation}
\label{eq12}
ds^2 = a^2(\tau) \left[-d\tau^2 + (\delta_{ij} + h_{ij}) dx^i dx^j \right], 
\end{equation}
where $\tau$ denotes the conformal time;  $\delta_{ij}$ and $h_{ij}$ are respectively the unperturbed and perturbed spatial part of the metric tensors. For the above metric, we can write the perturbation equations for all fluids present in the Hubble equation (\ref{eqn-Hubble}). Denoting $\delta_i = \delta \rho_i / \rho_i$ as the dimensionless density perturbations for the fluid component $i$ and $\theta_i = i \kappa^j v_j$ being the divergence of the $i$-th fluid velocity, their evolution equations in the Fourier space can be described as~\cite{Ma:1995ey}:
\begin{eqnarray}
\delta'_{i} & = & - (1+ w_{i})\, \left(\theta_{i}+ \frac{h'}{2}\right) - 
3\mathcal{H}\left(\frac{\delta P_i}{\delta \rho_i} - w_{i} \right)\delta_i \nonumber \\ 
& & -  9 \mathcal{H}^2\left(\frac{\delta P_i}{\delta \rho_i} - c^2_{a,i} \right) (1+w_i) 
\frac{\theta_i}
{{\kappa}^2}, \label{per1} \\
\theta'_{i} & = & - \mathcal{H} \left(1- 3 \frac{\delta P_i}{\delta
\rho_i}\right)\theta_{i} 
+ \frac{\delta P_i/\delta \rho_i}{1+w_{i}}\, {\kappa}^2\, \delta_{i} 
-{\kappa}^2\sigma_i,\label{per2}
\end{eqnarray}
where prime denotes the derivative with respect to the conformal time $\tau$;  $\mathcal{H}(a) = a'/a$ is the conformal Hubble parameter; $h$ is the usual synchronous gauge metric perturbation; $\kappa$ refers to the wavenumber in the Fourier space; $\sigma_i$ is the anisotropic stress of the $i$-th fluid,\footnote{We note that here anisotropic stress is neglected, i.e., $\sigma_i = 0$ has been set in the perturbation equations. } $ \frac{\delta P_{\rm DE}}{\delta \rho_{\rm DE}} = c^2_{\rm s,DE}$ is the square of the sound speed of the DE component in the rest frame, $c^2_{a,i} = w_i - w_i^{\prime}/(3 \mathcal{H} (1+w_i))$ is the adiabatic sound speed of the $i$-th fluid. In this article $c^2_{\rm s,DE}$ has been set to be unity similar to the minimally coupled scalar field models.

Using the evolution equations for background and perturbations, the complete evolution of the universe in the presence of a DDE fluid, characterized by its EoS $w(a)$, can be determined. 

In this review, we consider the well-known Chevallier-Polarski-Linder (CPL) parametrization~\cite{Chevallier:2000qy, Linder:2002et}, also employed by the DESI collaboration~\cite{DESI:2024mwx}, which is given by:

\begin{equation}
\label{cpl}
w(a) = w_0 + w_a \times \left(1 - a \right),
\end{equation}
where $w_0 = w(a_0)$ is the present-day value of $w(a)$, and $w_a = -dw(a)/da$ evaluated at $a_0 = 1$ determines the dynamical nature of the DE equation of state (EoS). For $w_a = 0$, the DE EoS reduces to a non-dynamical constant. 
From a theoretical standpoint, there is ongoing debate regarding the choice of this parametrization since Eq.~(\ref{cpl}) is merely a Taylor expansion of $w(a)$ around the present epoch, i.e., $a = 1$. Consequently, numerous alternative models have been proposed to describe a dynamical DE more accurately, often involving more complex but theoretically motivated functional forms for $w(a)$ (see, e.g., Refs.~\cite{Cooray:1999da,Efstathiou:1999tm,Wetterich:2004pv,Feng:2004ff,Xia:2004rw,Gong:2005de,Jassal:2005qc,Nesseris:2005ur,Liu:2008vy,Barboza:2008rh,Barboza:2009ks,Ma:2011nc,Sendra:2011pt,Feng:2011zzo,Barboza:2011gd,DeFelice:2012vd,Feng:2012gf,Wei:2013jya,Magana:2014voa,Akarsu:2015yea,Magana:2017usz,Pan:2017zoh,Panotopoulos:2018sso,Yang:2018qmz,Li:2019yem,Pan:2019hac,Tamayo:2019gqj,Pan:2019brc,Rezaei:2020mrj,Perkovic:2020mph,Banihashemi:2020wtb,Benaoum:2020qsi,Yang:2021eud,Escamilla:2023oce,Rezaei:2023xkj,Adil:2023exv,Giare:2024gpk,Rezaei:2024vtg,Wolf:2024eph,Wolf:2024stt,Wolf:2025jlc}). 
However, as evidenced in the literature, no widely accepted theoretical framework has yet emerged that can consistently align with all observational data. Therefore, given the ongoing complexities and the enigmatic nature of DE even after more than two decades since the discovery of cosmic acceleration, we avoid introducing additional theoretical complexity through convoluted forms of $w(a)$. Instead, we adopt this linear choice, i.e., Eq.~(\ref{cpl}), as a straightforward approach to investigate whether current observational data provide genuine evidence for a dynamical dark energy component. We close this section providing the evolution of the DE density for the CPL parametrization:

\begin{eqnarray}
    \rho_{\rm DE}/\rho_{\rm DE,0} = a^{-3 (1+w_0+w_a)} \times \exp\left[-3w_a (1-a) \right], 
\end{eqnarray}
from which one can sketch the qualitative feature of the DE evolution for different values of the DE parameters.

\section{Observational datasets and Methodology}
\label{sec:data+Methodology}

In this section, we present the observational datasets and the underlying methodology that we have used to constrain the proposed DE parametrizations. In what follows, we describe each dataset.

\begin{itemize}
    \item \textbf{CMB:} Planck 2018 (PR3) temperature and polarization (TT, TE, EE) power spectra~\cite{Planck:2019nip,Planck:2018vyg,Planck:2018nkj}, in combination with the Planck 2018 reconstructed lensing~\cite{Planck:2018lbu} and ACT-DR6 lensing~\cite{ACT:2023kun,ACT:2023dou} likelihoods have been used.
    
    \item \textbf{SDSS:} Baryon Acoustic Oscillations (BAO) and Redshift Space Distortions (RSD) measurements from the completed SDSS-IV eBOSS survey. These include isotropic and anisotropic distance and expansion rate measurements, as well as measurements of $f\sigma_8$~\cite{eBOSS:2020yzd}.
    
    \item \textbf{DESI:} Measurements of BAO in galaxy, quasar, and Lyman-$\alpha$ forest tracers from the first year of observations using DESI. This dataset also includes robust measurements of the Hubble rate and transverse comoving distance~\cite{DESI:2024uvr,DESI:2024lzq,DESI:2024mwx}.
    
    \item \textbf{PantheonPlus (PP):} We use the 1701 light curves of 1550 distinct Type Ia supernovae (SNIa), including those with $z<0.01$, to allow for a joint measurement of the DE EoS parameter and the Hubble constant $H_0$~\cite{Scolnic:2021amr,Brout:2022vxf}.
    
    \item \textbf{DESY5:} We have also considered the distance modulus measurements of 1635 Type Ia SN distributed in $0.10 < z < 1.13$ by the Dark Energy Survey (DES) Supernova Program during five years~\cite{DES:2024tys,DES:2024upw,DES:2024hip}, together with 194 low-redshift SNIa in $0.025 < z < 0.1$ that overlap with the PantheonPlus sample~\cite{Scolnic:2021amr,Brout:2022vxf}.
    
    \item \textbf{Union3 (U3):} We use the compilation of 2087 Type Ia SN, of which 1363 SNIa in common with the PantheonPlus sample, as presented in Ref.~\cite{Rubin:2023ovl}, that includes improvements in the model selection effects, standardization, and systematic uncertainties.
    

    \item \textbf{Cosmic Chronometers (CC):} The expansion rate $H(z)$ is determined from the relative ages of massive, passively evolving galaxies at early cosmic times, known as Cosmic Chronometers~\cite{Jimenez:2001gg}. For our analysis, we adopt a conservative approach, utilizing a dataset of 15 CC measurements within the redshift range $z \sim 0.179$ to $z \sim 1.965$~\cite{Moresco:2012jh,Moresco:2015cya,Moresco:2016mzx}, incorporating all non-diagonal elements of the covariance matrix and accounting for systematic uncertainties.
\end{itemize}

For the statistical analyses, we have used the publicly available cosmological package \texttt{CAMB}~\cite{Lewis:1999bs,Lewis:2002ah,Howlett:2012mh} and the sampler \texttt{Cobaya}~\cite{Lewis:2013hha} to perform the Markov Chain Monte Carlo (MCMC) analyses. The convergence of the produced MCMC chains was assessed using the Gelman-Rubin statistic, $R-1$~\cite{Gelman:1992zz}. The MCMC chains were run until $R-1 < 0.02$. 
In Table~\ref{tab-priors}, we list the flat priors applied to the free parameters in the current model. The descriptions of the free parameters are as follows: $\Omega_\mathrm{b} h^2$ is the physical baryon energy density, $\Omega_\mathrm{c} h^2$ is the physical cold dark matter energy density, $A_\mathrm{s}$ denotes the amplitude of the primordial scalar spectrum, $n_s$ is the spectral index of the primordial scalar spectrum, $\tau$ corresponds to the optical depth to reionization, $\theta_{\rm{MC}}$ is the angular size of the sound horizon, and there are two additional free parameters, $w_0$ and $w_a$, originating from the CPL parametrization.

We assess the preference for an evolving DE equation of state by computing the difference in the minimum $\chi^2$ obtained assuming a CPL parametrization for DE and a standard $\Lambda$CDM cosmology. In particular, assuming both models are analyzed with the same dataset, the difference $\Delta \chi^2 = \min(\chi^2_{\rm CPL}) - \min(\chi^2_{\Lambda\rm{CDM}})$ follows approximately a $\chi^2$ distribution with degrees of freedom equal to the difference in the number of free parameters between the models, $\Delta k = k_{\rm CPL} - k_{\Lambda\rm{CDM}}=2$. The probability of obtaining a $\Delta \chi^2$ as extreme as the observed value, under the null hypothesis that the models provide an equally good fit to the data, is given by the $p$-value:
\begin{equation}
p = 1 - F_{\chi^2}(|\Delta \chi^2|, \Delta k),
\end{equation}
where $F_{\chi^2}(x, k)$ is the cumulative distribution function of the $\chi^2$ distribution with $k$ degrees of freedom:
\begin{equation}
F_{\chi^2}(x, k) = \frac{1}{2^{k / 2} \Gamma(k / 2)} \int_{0}^{x} t^{k / 2 - 1} e^{-t/2} dt.
\end{equation}
Ultimately, we convert the $p$-value into a tension in units of standard deviations (\#$\sigma$) 
\begin{equation}
\sigma = \Phi^{-1}(1 - p/2),
\end{equation}
by using the (inverse of the) cumulative distribution function of the standard normal distribution, $\Phi$ defined as:
\begin{equation}
\Phi(x) = \frac{1}{\sqrt{2\pi}} \int_{-\infty}^x e^{-t^2 / 2} dt.
\end{equation}

\begin{table}
\centering
\renewcommand{\arraystretch}{1.5}
\begin{tabular}{l @{\hspace{2cm}} c}
\toprule
\textbf{Parameter} & \textbf{Prior} \\
\hline \hline
$\Omega_\mathrm{b} h^2$ & $[0.005, 0.1]$ \\
$\Omega_\mathrm{c} h^2$ & $[0.01, 0.99]$ \\
$\log(10^{10} A_\mathrm{s})$ & $[1.61, 3.91]$ \\
$n_\mathrm{s}$ & $[0.8, 1.2]$ \\
$\tau$ & $[0.01, 0.8]$ \\
$100\theta_\mathrm{MC}$ & $[0.5, 10]$ \\
$w_0$ & $[-2, 0]$ \\
$w_a$ & $[-2, 2]$ \\
\bottomrule
\end{tabular}
\caption{Flat priors on the cosmological parameters sampled in this work.}
\label{tab-priors}
\end{table}

\begin{table*}[htp!]
\footnotesize
\centering 
\renewcommand{\arraystretch}{2.1}
\resizebox{0.95 \textwidth}{!}{
\begin{tabular}{lccccr} 
\hline 
\textbf{Dataset} & \boldmath{$w_{0}$} & \boldmath{$w_a$} & \boldmath{$H_0\,[{\rm km}/{\rm s}/{\rm Mpc}]$} & \boldmath{$\Omega_m$} & \boldmath{\# $\sigma$} \\ 
\hline 
\textbf{ CMB } &$  -1.36^{+0.24}_{-0.53}\, (< -0.619 ) $ &$ < -0.197\, (< 0.967 ) $ &$  84^{+20}_{-5}\, (> 66.5 ) $ &$  0.209^{+0.017}_{-0.066}\, ( 0.209^{+0.12}_{-0.070} ) $ &$ 1.9 $\\ 

\textbf{ CMB+CC } &$  -1.03^{+0.48}_{-0.37}\, ( -1.03^{+0.70}_{-0.80} ) $ &$  -0.47^{+0.54}_{-1.4}\, (< 1.13 ) $ &$  72.3^{+5.7}_{-7.7}\, ( 72^{+10}_{-10} ) $ &$  0.280^{+0.045}_{-0.059}\, ( 0.280^{+0.10}_{-0.097} ) $ &$ 0.1 $\\ 

\textbf{ CMB+DESI } &$  -0.63^{+0.26}_{-0.14}\, ( -0.63^{+0.35}_{-0.43} ) $ &$ < -1.05\, (< -0.238 ) $ &$  66.3^{+1.9}_{-2.8}\, ( 66.3^{+5.1}_{-4.5} ) $ &$  0.326^{+0.026}_{-0.021}\, ( 0.326^{+0.045}_{-0.048} ) $ &$ 2.3 $\\ 

\textbf{ CMB+DESI+CC } &$  -0.65^{+0.26}_{-0.14}\, ( -0.65^{+0.35}_{-0.42} ) $ &$ < -0.991\, (< -0.226 ) $ &$  66.3^{+1.8}_{-2.8}\, ( 66.3^{+4.9}_{-4.4} ) $ &$  0.326^{+0.026}_{-0.021}\, ( 0.326^{+0.044}_{-0.047} ) $ &$ 2.6 $\\ 

\textbf{ CMB+DESI+DESY5 } &$  -0.735\pm 0.067\, ( -0.73^{+0.13}_{-0.13} ) $ &$  -0.999^{+0.31}_{-0.28}\, ( -0.999^{+0.56}_{-0.62} ) $ &$  67.27\pm 0.66\, ( 67.3^{+1.3}_{-1.3} ) $ &$  0.3156\pm 0.0065\, ( 0.316^{+0.013}_{-0.013} ) $ &$ 3.9 $\\ 

\textbf{ CMB+DESI+DESY5+CC } &$  -0.737\pm 0.065\, ( -0.74^{+0.13}_{-0.13} ) $ &$  -0.99^{+0.30}_{-0.27}\, ( -0.99^{+0.54}_{-0.62} ) $ &$  67.23\pm 0.65\, ( 67.2^{+1.2}_{-1.3} ) $ &$  0.3161\pm 0.0064\, ( 0.316^{+0.013}_{-0.012} ) $ &$ 3.8 $\\ 

\textbf{ CMB+DESI+PP } &$  -0.824\pm 0.064\, ( -0.82^{+0.13}_{-0.12} ) $ &$  -0.74^{+0.29}_{-0.25}\, ( -0.74^{+0.50}_{-0.58} ) $ &$  67.99\pm 0.71\, ( 68.0^{+1.4}_{-1.4} ) $ &$  0.3089\pm 0.0067\, ( 0.309^{+0.013}_{-0.013} ) $ &$ 2.5 $\\ 

\textbf{ CMB+DESI+PP+CC } &$  -0.827\pm 0.061\, ( -0.83^{+0.13}_{-0.12} ) $ &$  -0.72^{+0.28}_{-0.24}\, ( -0.72^{+0.49}_{-0.55} ) $ &$  67.96\pm 0.71\, ( 68.0^{+1.4}_{-1.4} ) $ &$  0.3091\pm 0.0068\, ( 0.309^{+0.014}_{-0.013} ) $ &$ 2.5 $\\ 

\textbf{ CMB+DESI+U3 } &$  -0.661\pm 0.095\, ( -0.66^{+0.18}_{-0.19} ) $ &$  -1.20\pm 0.34\, ( -1.20^{+0.63}_{-0.71} ) $ &$  66.58\pm 0.94\, ( 66.6^{+1.8}_{-1.8} ) $ &$  0.3225\pm 0.0094\, ( 0.322^{+0.019}_{-0.018} ) $ &$ 3.5 $\\ 

\textbf{ CMB+DESI+U3+CC } &$  -0.669\pm 0.093\, ( -0.67^{+0.18}_{-0.19} ) $ &$  -1.16^{+0.36}_{-0.33}\, ( -1.16^{+0.64}_{-0.67} ) $ &$  66.57\pm 0.95\, ( 66.6^{+1.9}_{-1.8} ) $ &$  0.3224\pm 0.0095\, ( 0.322^{+0.019}_{-0.018} ) $ &$ 3.3 $\\ 

\textbf{ CMB+DESY5 } &$  -0.737^{+0.10}_{-0.089}\, ( -0.74^{+0.18}_{-0.18} ) $ &$  -1.02\pm 0.46\, ( -1.02^{+0.83}_{-0.90} ) $ &$  67.37\pm 0.98\, ( 67.4^{+1.9}_{-2.0} ) $ &$  0.3150^{+0.0094}_{-0.011}\, ( 0.315^{+0.021}_{-0.019} ) $ &$ 2.6 $\\ 

\textbf{ CMB+PP } &$  -0.860^{+0.12}_{-0.093}\, ( -0.86^{+0.20}_{-0.22} ) $ &$  -0.61^{+0.51}_{-0.60}\, ( -0.6^{+1.0}_{-1.0} ) $ &$  67.9\pm 1.3\, ( 67.9^{+2.4}_{-2.5} ) $ &$  0.310^{+0.013}_{-0.015}\, ( 0.310^{+0.026}_{-0.024} ) $ &$ 0.1 $\\ 

\textbf{ CMB+SDSS } &$  -0.80\pm 0.19\, ( -0.80^{+0.37}_{-0.37} ) $ &$  -0.68^{+0.58}_{-0.51}\, ( -0.68^{+0.99}_{-1.1} ) $ &$  66.8^{+1.8}_{-2.1}\, ( 66.8^{+3.9}_{-3.6} ) $ &$  0.322\pm 0.019\, ( 0.322^{+0.037}_{-0.036} ) $ &$ 0.3 $\\ 

\textbf{ CMB+SDSS+CC } &$  -0.83\pm 0.19\, ( -0.83^{+0.36}_{-0.36} ) $ &$  -0.59^{+0.56}_{-0.49}\, ( -0.59^{+0.96}_{-1.1} ) $ &$  67.0^{+1.7}_{-2.0}\, ( 67.0^{+3.9}_{-3.5} ) $ &$  0.320\pm 0.018\, ( 0.320^{+0.036}_{-0.035} ) $ &$ 0.3 $\\ 

\textbf{ CMB+SDSS+DESY5 } &$  -0.800\pm 0.064\, ( -0.80^{+0.13}_{-0.12} ) $ &$  -0.69^{+0.28}_{-0.24}\, ( -0.69^{+0.49}_{-0.55} ) $ &$  66.75\pm 0.62\, ( 66.8^{+1.2}_{-1.2} ) $ &$  0.3215\pm 0.0064\, ( 0.322^{+0.013}_{-0.012} ) $ &$ 2.6 $\\ 

\textbf{ CMB+SDSS+DESY5+CC } &$  -0.804\pm 0.063\, ( -0.80^{+0.13}_{-0.12} ) $ &$  -0.67^{+0.28}_{-0.23}\, ( -0.67^{+0.48}_{-0.54} ) $ &$  66.71\pm 0.62\, ( 66.7^{+1.2}_{-1.2} ) $ &$  0.3219\pm 0.0065\, ( 0.322^{+0.013}_{-0.013} ) $ &$ 2.3 $\\ 

\textbf{ CMB+SDSS+PP } &$  -0.879\pm 0.060\, ( -0.88^{+0.12}_{-0.12} ) $ &$  -0.46^{+0.25}_{-0.22}\, ( -0.46^{+0.45}_{-0.50} ) $ &$  67.37\pm 0.68\, ( 67.4^{+1.3}_{-1.3} ) $ &$  0.3154\pm 0.0068\, ( 0.315^{+0.013}_{-0.013} ) $ &$ 1.6 $\\ 

\textbf{ CMB+SDSS+PP+CC } &$  -0.885\pm 0.061\, ( -0.89^{+0.12}_{-0.12} ) $ &$  -0.43^{+0.25}_{-0.22}\, ( -0.43^{+0.44}_{-0.49} ) $ &$  67.33\pm 0.66\, ( 67.3^{+1.3}_{-1.3} ) $ &$  0.3158\pm 0.0066\, ( 0.316^{+0.013}_{-0.013} ) $ &$ 1.0 $\\ 

\textbf{ CMB+SDSS+U3 } &$  -0.751\pm 0.093\, ( -0.75^{+0.18}_{-0.18} ) $ &$  -0.82^{+0.35}_{-0.31}\, ( -0.82^{+0.62}_{-0.67} ) $ &$  66.26\pm 0.87\, ( 66.3^{+1.7}_{-1.7} ) $ &$  0.3264\pm 0.0089\, ( 0.326^{+0.018}_{-0.017} ) $ &$ 1.9 $\\ 

\textbf{ CMB+SDSS+U3+CC } &$  -0.761\pm 0.094\, ( -0.76^{+0.19}_{-0.18} ) $ &$  -0.78^{+0.35}_{-0.31}\, ( -0.78^{+0.61}_{-0.67} ) $ &$  66.30\pm 0.87\, ( 66.3^{+1.7}_{-1.7} ) $ &$  0.3260\pm 0.0089\, ( 0.326^{+0.018}_{-0.018} ) $ &$ 1.8 $\\ 

\textbf{ CMB+U3 } &$  -0.677^{+0.13}_{-0.096}\, ( -0.68^{+0.20}_{-0.23} ) $ &$  -1.19^{+0.33}_{-0.69}\, (< -0.302 ) $ &$  66.9\pm 1.3\, ( 66.9^{+2.5}_{-2.5} ) $ &$  0.319\pm 0.013\, ( 0.319^{+0.027}_{-0.024} ) $ &$ 1.7 $\\ 

\hline

\textbf{ DESI+CC } &$  -0.80^{+0.30}_{-0.23}\, ( -0.80^{+0.47}_{-0.50} ) $ &$ < -0.197\, (< 0.947 ) $ &$  67.8^{+4.2}_{-4.9}\, ( 68^{+9}_{-9} ) $ &$  0.313^{+0.048}_{-0.028}\, ( 0.313^{+0.065}_{-0.081} ) $ &$ 0.9 $\\ 

\textbf{ DESI+DESY5 } &$  -0.755\pm 0.079\, ( -0.75^{+0.15}_{-0.15} ) $ &$  -1.00^{+0.43}_{-0.74}\, (< 0.0676 ) $ &$ > 78.9\, (> 62.5 ) $ &$  0.323^{+0.022}_{-0.013}\, ( 0.323^{+0.040}_{-0.041} ) $ &$ 2.9 $\\ 

\textbf{ DESI+DESY5+CC } &$  -0.763\pm 0.077\, ( -0.76^{+0.15}_{-0.15} ) $ &$  -0.98^{+0.49}_{-0.67}\, (< -0.0347 ) $ &$  67.2\pm 3.7\, ( 67.2^{+7.1}_{-7.0} ) $ &$  0.325^{+0.019}_{-0.015}\, ( 0.325^{+0.034}_{-0.036} ) $ &$ 2.8 $\\ 

\textbf{ DESI+PP } &$  -0.870^{+0.065}_{-0.073}\, ( -0.87^{+0.14}_{-0.13} ) $ &$  -0.31^{+0.72}_{-0.65}\, ( -0.3^{+1.2}_{-1.2} ) $ &$ > 74.8\, (> 56.3 ) $ &$  0.299^{+0.036}_{-0.013}\, ( 0.299^{+0.049}_{-0.072} ) $ &$ 1.5 $\\ 

\textbf{ DESI+PP+CC } &$  -0.871^{+0.066}_{-0.078}\, ( -0.87^{+0.15}_{-0.14} ) $ &$  -0.43^{+0.66}_{-0.54}\, ( -0.4^{+1.1}_{-1.1} ) $ &$  68.2\pm 3.7\, ( 68.2^{+7.4}_{-7.2} ) $ &$  0.307^{+0.025}_{-0.016}\, ( 0.307^{+0.042}_{-0.048} ) $ &$ 0.0 $\\ 

\textbf{ DESI+U3 } &$  -0.701^{+0.11}_{-0.092}\, ( -0.70^{+0.18}_{-0.21} ) $ &$  -1.09^{+0.28}_{-0.85}\, (< 0.0391 ) $ &$ > 78.8\, (> 63.0 ) $ &$  0.327^{+0.023}_{-0.014}\, ( 0.327^{+0.041}_{-0.045} ) $ &$ 2.5 $\\ 

\textbf{ DESI+U3+CC } &$  -0.70^{+0.12}_{-0.10}\, ( -0.70^{+0.20}_{-0.21} ) $ &$ < -0.906\, (< -0.107 ) $ &$  66.9\pm 3.7\, ( 66.9^{+7.2}_{-6.9} ) $ &$  0.330^{+0.021}_{-0.015}\, ( 0.330^{+0.035}_{-0.040} ) $ &$ 2.4 $\\ 

\textbf{ SDSS+CC } &$  -0.90\pm 0.24\, ( -0.90^{+0.47}_{-0.45} ) $ &$  -0.09^{+0.92}_{-0.59}\, ( -0.1^{+1.3}_{-1.5} ) $ &$  68.3\pm 4.9\, ( 68^{+10}_{-9} ) $ &$  0.297^{+0.028}_{-0.035}\, ( 0.297^{+0.061}_{-0.057} ) $ &$ 0.7 $\\ 

\textbf{ SDSS+DESY5 } &$  -0.805^{+0.065}_{-0.084}\, ( -0.80^{+0.16}_{-0.14} ) $ &$  -0.46^{+0.70}_{-0.46}\, ( -0.5^{+1.0}_{-1.2} ) $ &$ > 80.0\, (> 61.6 ) $ &$  0.307^{+0.027}_{-0.022}\, ( 0.307^{+0.042}_{-0.047} ) $ &$ 2.5 $\\ 

\textbf{ SDSS+DESY5+CC } &$  -0.811^{+0.066}_{-0.079}\, ( -0.81^{+0.15}_{-0.13} ) $ &$  -0.45^{+0.66}_{-0.38}\, ( -0.45^{+0.90}_{-1.1} ) $ &$  67.3\pm 3.8\, ( 67^{+8}_{-7} ) $ &$  0.309\pm 0.020\, ( 0.309^{+0.040}_{-0.038} ) $ &$ 2.5 $\\ 

\textbf{ SDSS+PP } &$  -0.897\pm 0.062\, ( -0.90^{+0.13}_{-0.12} ) $ &$  0.01^{+0.54}_{-0.29}\, ( 0.01^{+0.78}_{-0.97} ) $ &$ > 81.3\, (> 64.4 ) $ &$  0.291^{+0.022}_{-0.024}\, ( 0.291^{+0.043}_{-0.042} ) $ &$ 1.6 $\\ 

\textbf{ SDSS+PP+CC } &$  -0.901\pm 0.062\, ( -0.90^{+0.13}_{-0.12} ) $ &$  -0.03^{+0.47}_{-0.26}\, ( -0.03^{+0.69}_{-0.84} ) $ &$  68.4\pm 3.9\, ( 68.4^{+7.5}_{-7.5} ) $ &$  0.295^{+0.017}_{-0.021}\, ( 0.295^{+0.037}_{-0.035} ) $ &$ 1.6 $\\ 

\textbf{ SDSS+U3 } &$  -0.763^{+0.091}_{-0.11}\, ( -0.76^{+0.21}_{-0.19} ) $ &$  -0.49^{+0.75}_{-0.45}\, ( -0.49^{+0.99}_{-1.2} ) $ &$ > 79.1\, (> 61.3 ) $ &$  0.309\pm 0.024\, ( 0.309^{+0.046}_{-0.048} ) $ &$ 2.2 $\\ 

\textbf{ SDSS+U3+CC } &$  -0.777^{+0.097}_{-0.11}\, ( -0.78^{+0.20}_{-0.19} ) $ &$  -0.51^{+0.72}_{-0.45}\, ( -0.51^{+0.98}_{-1.2} ) $ &$  66.8\pm 4.2\, ( 67^{+8}_{-8} ) $ &$  0.312\pm 0.022\, ( 0.312^{+0.043}_{-0.041} ) $ &$ 2.1 $\\

\hline\hline 

\end{tabular}}
\caption{Summary of the constraints on key parameters (i.e.,  the dark energy parameters, $w_0$ and $w_a$, the Hubble constant $H_0$, and the present matter density parameter $\Omega_m$) at 68\% CL (95\% CL) for various dataset combinations. For cases where parameters are unconstrained, we report their upper or lower limits. The upper half of the table presents constraints based on high-redshift CMB data, both alone and in combination with BAO measurements (from either DESI or SDSS), SN distance moduli (from PP, DESY5, and U3), and CC. The lower half shows constraints obtained exclusively from low-redshift probes. In the last column, we quantify the preference for DDE in terms of the standard deviation $\sigma$, computed as described in Sec.~\ref{sec:data+Methodology}.}
\label{table:CPL-w0wa}
\end{table*}


\begin{figure*}
    \centering
    \includegraphics[width=1\textwidth]{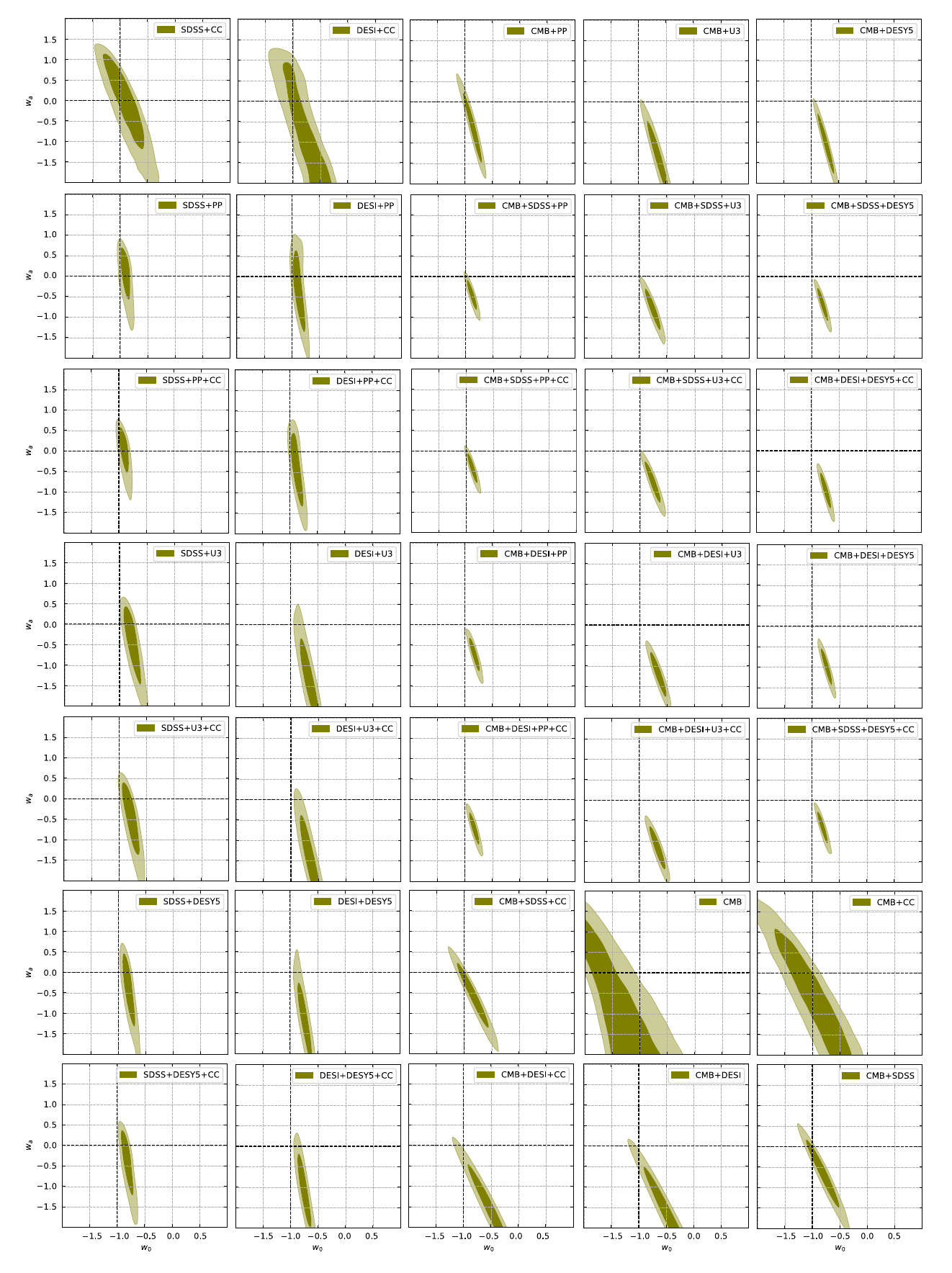}
    \caption{The $w_0$-$w_a$ plane for the CPL parametrization at 68\% and 95\% CL for all datasets analyzed in this article.}
    \label{fig:2D-w0wa}
\end{figure*}
\begin{figure*}
    \centering
    \includegraphics[width=1\textwidth]{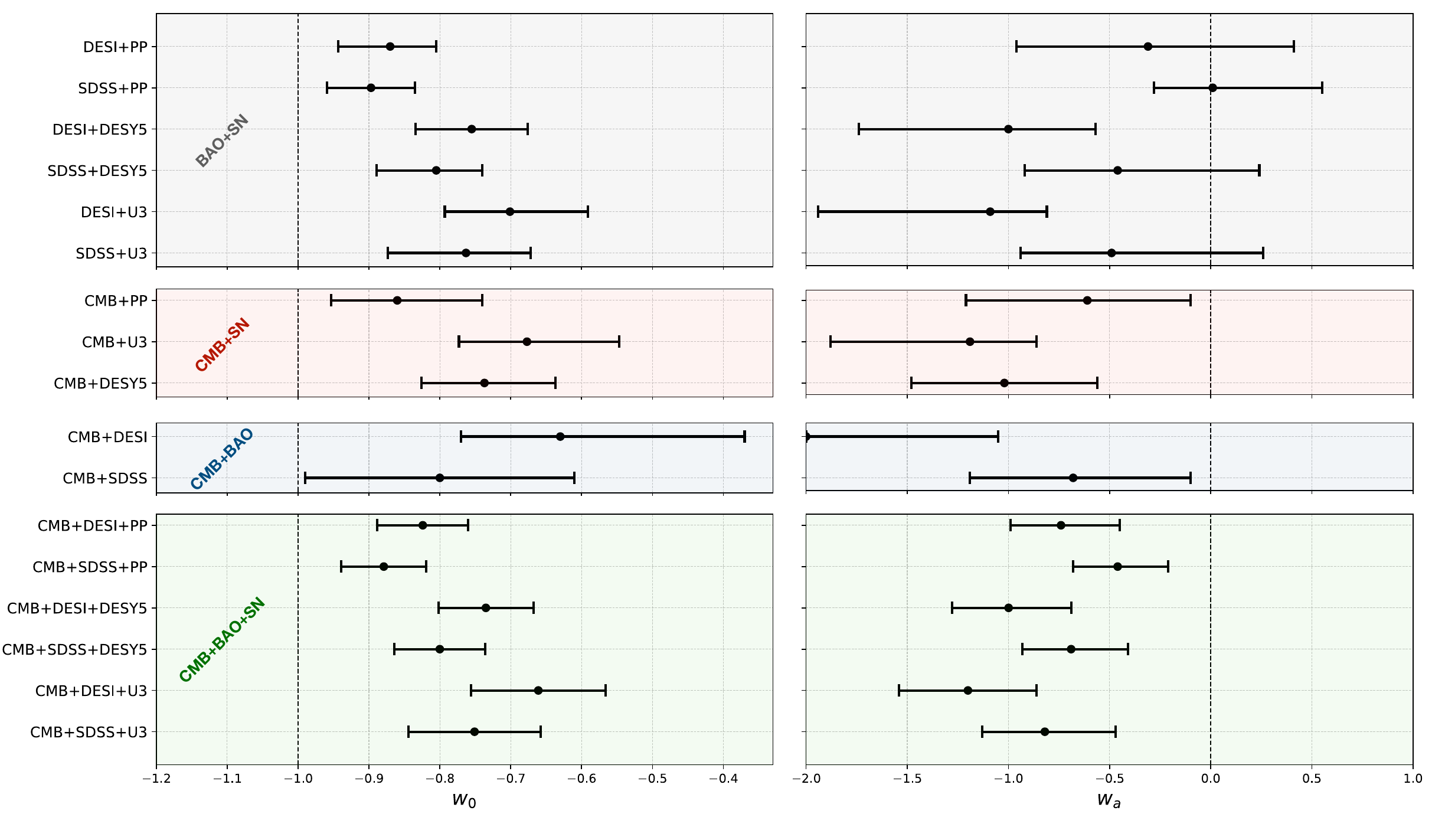}
    \caption{Whisker plots with the 68\% CL constraints on the DE parameters, $w_0$ (left plot) and $w_a$ (right plot) for a wide variety of combined cosmological probes. The inclusion of CC does not significancy affect the constraints on $w_0$ and $w_a$, therefore we do not consider additional cases incorporating CC.} 
    \label{fig:whisker-w0wa}
\end{figure*}
\begin{figure*}
    \centering
    \includegraphics[width=1\textwidth]{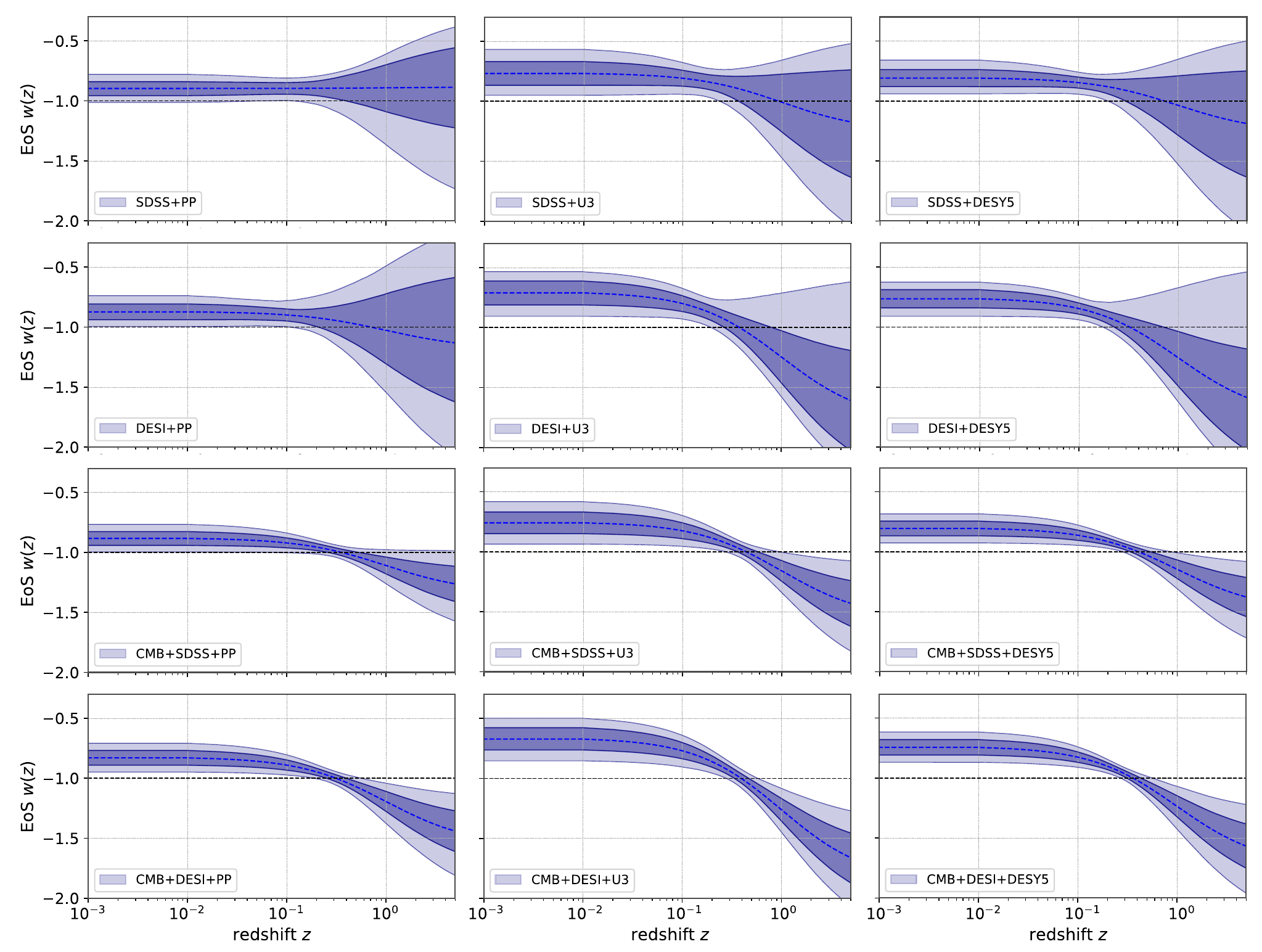}
    \caption{Evolution of the CPL EoS $w(z)$ considering various combined astronomical probes. The blue dashed lines correspond to the mean values of $w(z)$ and the solid lines are denoting its $1\sigma$ and $2\sigma$ uncertainties. }
    \label{fig:eos-w}
\end{figure*}
\begin{figure}
    \centering
    \includegraphics[width=1\linewidth]{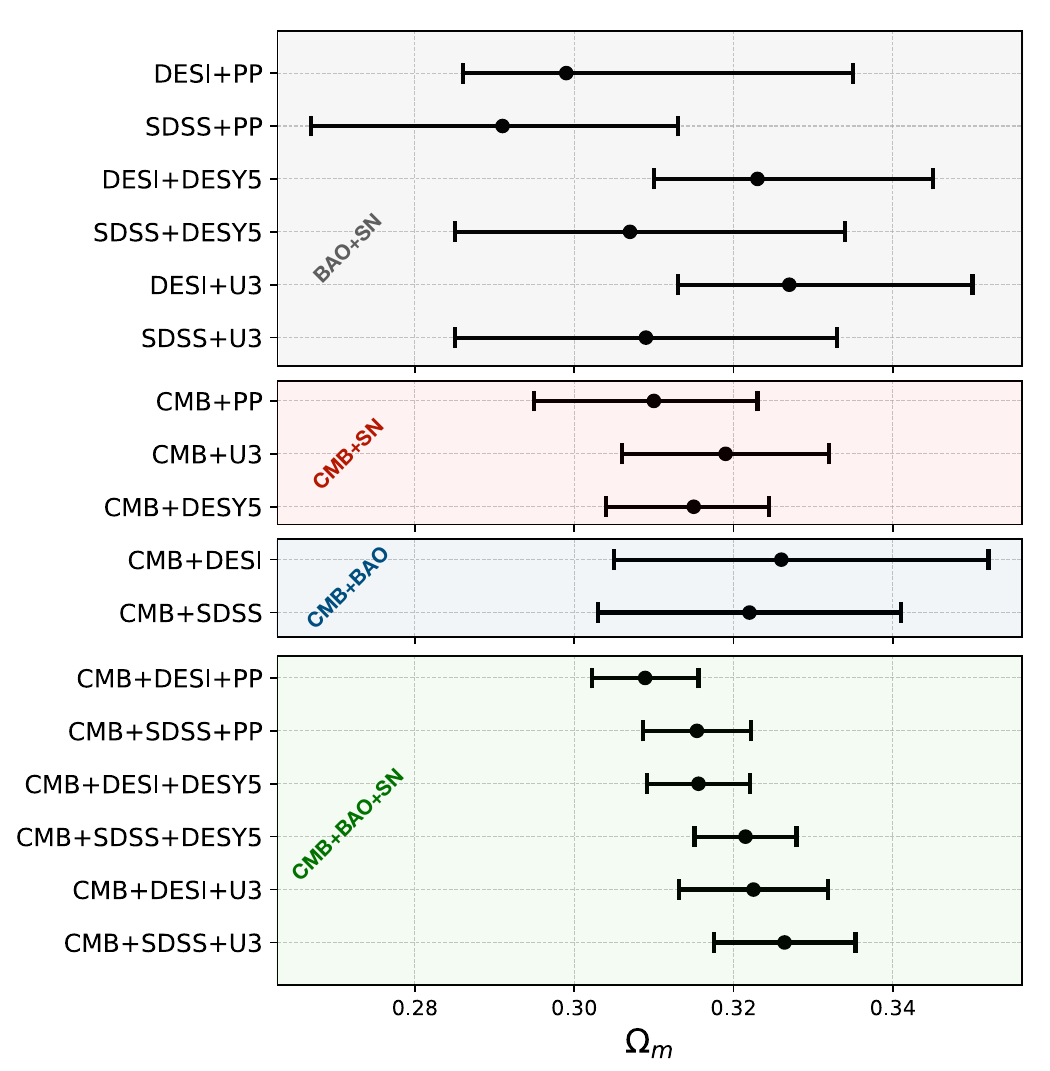}
    \caption{Whisker plot showing the 68\% CL constraints on the present-day matter density parameter, $\Omega_m$, for various combinations of cosmological probes. The inclusion of CC does not significancy impact the results on $\Omega_m$; therefore, we do not consider additional cases incorporating CC.} 
    \label{fig:whisker-Omegam}
\end{figure}

\section{Results}
\label{sec:Results}

We present the constraints on the CPL parametrization by considering several cosmological probes. We perform two separate analyses with this model. First, we use only the low-redshift datasets (SN, BAO, CC). Then, we incorporate CMB data from Planck 2018 \cite{Planck:2018vyg} alongside all the low-redshift cosmological probes. This distinction allows us to explore the differences in the constraints derived from the low redshift probes and CMB. In Table~\ref{table:CPL-w0wa}, we summarize the constraints on the key parameters of interest, namely the dark energy parameters, $w_0$ and $w_a$, the Hubble constant $H_0$, and the present matter density parameter $\Omega_m$.\footnote{For discussions on this parameter see Refs.~\cite{Lin:2019htv,Jedamzik:2020zmd,Lin:2021sfs,Baryakhtar:2024rky,Poulin:2024ken,Pedrotti:2024kpn}.} The first half of Table~\ref{table:CPL-w0wa} presents the constraints on these model parameters for CMB alone and when CMB is included with the low redshift probes, while the second half focuses on results derived solely from the low-redshift cosmological probes. The graphical summary representing  this model are shown in Figs. \ref{fig:2D-w0wa}, \ref{fig:whisker-w0wa}, \ref{fig:eos-w}, \ref{fig:whisker-Omegam}, \ref{fig:sigma}. In the following, we systematically discuss the results obtained from all these dataset combinations, while postponing a summary and key highlights of the main findings to the conclusions.

\subsection{Constraints from the low-redshift probes}

Here, we present the constraints on the key parameters of the model obtained from various combinations of low-redshift datasets, specifically BAO from two distinct surveys, namely, DESI, SDSS; CC measurements, and three distinct samples of SN, namely, DESY5, PP, and U3.\footnote{According to Refs.~\cite{Efstathiou:2024xcq,Huang:2025som}, the DESY5 sample might have some systematics that could affect the evidence for DDE (see also Ref.~\cite{Dhawan:2024gqy}). Although such possibility may not be excluded, in this work we present the results combining DESY5 with other datasets as well without entering into this debate.}

\subsubsection{BAO+CC}

We include CC with two versions of the BAO data, and specifically performe the analyses with DESI+CC and SDSS+CC. According to the results, we note that only for the SDSS+CC dataset, $w_a$ is constrained, while DESI+CC fails to constrain it within the considered flat prior, i.e., $w_a \in [-2, 2]$. In both cases, no significant evidence for a non-zero $w_a$ is found. Furthermore, while the mean value of $w_0$ remains in the quintessential regime, within $1\sigma$, the value $w_0 = -1$ is still consistent with the data. Consequently, neither dataset has enough constraining power to provide positive evidence for a dynamical evolution of $w(a)$.

On the other hand, the mean values of $H_0$ and $\Omega_{\rm m}$ for DESI+CC are almost identical to the Planck 2018 estimations (assuming the $\Lambda$CDM model in the background)~\cite{Planck:2018vyg}, though they exhibit significantly larger error bars. This can be visualized by comparing DESI+CC constraints ($\Omega_{\rm m} = 0.313^{+0.048}_{-0.028}$ at 68\% CL, $H_0 = 67.8^{+4.2}_{-4.9}$~km/s/Mpc at 68\% CL) with those from Planck 2018 TT,TE,EE+lowE+lensing~\cite{Planck:2018vyg} ($\Omega_{\rm m} = 0.3153 \pm 0.0073$ at 68\% CL, $H_0 = 67.36 \pm 0.54$~km/s/Mpc at 68\% CL). 
However, the SDSS+CC combination leads to a slightly higher mean value of $H_0$ ($H_0 = 68.3 \pm 4.9$~km/s/Mpc at 68\% CL) compared to both DESI+CC and Planck 2018 results. Consequently, this results in a mildly lower value of $\Omega_{\rm m}$. Nonetheless, the error bars on both $H_0$ and $\Omega_{\rm m}$ are similarly large as those observed for DESI+CC. 
Overall, neither DESI+CC nor SDSS+CC suggests any significant deviation from the standard $\Lambda$CDM model, as statistically, the CPL model remains within $1\sigma$ ($0.9\sigma$ and $0.7\sigma$ for DESI+CC and SDSS+CC, respectively; see the last column of Table~\ref{table:CPL-w0wa}).

\subsubsection{BAO+SN}

We consider six distinct combinations of data, involving two different versions of BAO (DESI and SDSS) and three different samples of SN (DESY5, PP, U3). Additionally, the same cases have been tested with the inclusion of CC data.

From the plots in Fig.~\ref{fig:2D-w0wa}, in the first two columns, it is evident that the cosmological constant is ruled out at more than 95\% CL for all the cases where U3 or DESY5 is included. Conversely, the indication is reduced to 68\% CL when PP is considered. A quantification of the deviation from $\Lambda$CDM can be found in the last column of Table~\ref{table:CPL-w0wa}, where it is confirmed that in the cases involving U3 or DESY5, the sole combination of BAO+SN is sufficient to provide evidence for a DDE at more than $2\sigma$, which can be visualized in the bottom of the radar plot in Fig.~\ref{fig:sigma}.

Focusing separately on the results for $w_0$ and $w_a$ from Fig.~\ref{fig:whisker-w0wa}, the non-null nature of $w_a$ is detected for DESI+DESY5 and DESI+U3 at more than $1\sigma$, while for the other cases, it remains consistent with zero. On the other hand, a quintessential nature of $w(a)$ at the present time, with  $w_0 > -1$, is identified for all dataset combinations at least at 68\% CL, reaching more than $2\sigma$ when U3 or DESY5 is included (see also Fig.~\ref{fig:eos-w} for the evolution of $w(a)$). It might be interesting to note that Fig.~\ref{fig:eos-w} (see the first two rows) clearly  depicts the transition (except for SDSS+PP where this transition is not present within this redshift region)\footnote{However, as we will discuss later, when CMB is added to SDSS+PP, the transition becomes then clear.}  of $w(a)$ from its past phantom-like nature ($w(a) < -1$) to the quintessential nature at current times.   
The matter density parameter remains stable and almost consistent with the $\Lambda$CDM expectations,\footnote{Although, upon a critical examination of the predicted mean values of the matter density parameter, one can conclude that the DESI+DESY5 and DESI+U3 datasets favor a slightly higher value of $\Omega_{\rm m}$ compared to Planck 2018 TT,TE,EE+lowE+lensing~\cite{Planck:2018vyg}, while DESI+PP predicts a lower value. On the contrary, the SDSS+SN datasets tend to prefer lower mean values of $\Omega_{\rm m}$ across all cases. } as shown in Fig.~\ref{fig:whisker-Omegam}, while the value of the Hubble constant is consistent with the local measurements obtained by SH0ES, primarily due to the large error bars.

We also examine possible improvements resulting from the inclusion of CC. Focusing first on the DESI+SN+CC datasets, and referring to the second column of Fig.~\ref{fig:2D-w0wa}, the cosmological constant is excluded at more than $2\sigma$ when DESY5 or U3 is included, while this evidence is reduced to approximately $1\sigma$ when PP is part of the combined dataset. This clearly indicates that the inclusion of CC does not significantly alter the evidence for DDE, as previously observed in the DESI+SN datasets.
In particular, the nature of $w_0$ remains unchanged with the addition of CC, with $w_0 > -1$ at more than $2\sigma$ when DESY5 or U3 is included, while for PP, the evidence weakens to between $1\sigma$ and $2\sigma$. Regarding $w_a$, no significant changes arise after the inclusion of CC for DESY5 ($w_a$ remains non-null within $1\sigma$) or PP (where $w_a = 0$ is allowed within $1\sigma$). However, the indication for $w_a \neq 0$ at more than $1\sigma$ observed for DESI+U3 disappears when CC is included, i.e., for DESI+U3+CC. 
On the other hand, the inclusion of CC shifts the estimation of $H_0$ closer to the Planck 2018 results~\cite{Planck:2018vyg}, though with larger error bars. This behavior is a direct effect introduced by the CC data. Specifically, the mean value of $H_0$ for DESI+DESY5+CC is quite close to Planck's mean value (Planck 2018 TT,TE,EE+lowE+lensing)~\cite{Planck:2018vyg}, but it increases slightly for DESI+PP+CC and decreases for DESI+U3+CC. 
Since $\Omega_{\rm m}$ and $H_0$ are anti-correlated, $\Omega_{\rm m}$ adjusts proportionally to variations in $H_0$. For instance, the lowest estimated value of $H_0$ ($H_0 = 66.9 \pm 3.7$ km/s/Mpc at 68\% CL) for DESI+U3+CC corresponds to a higher matter density $\Omega_{\rm m} = 0.330^{+0.021}_{-0.015}$ at 68\% CL. Conversely, the highest estimation of $H_0$ ($H_0 = 68.2 \pm 3.7$ km/s/Mpc at 68\% CL) obtained for DESI+PP+CC leads to a lower value of $\Omega_{\rm m} = 0.307^{+0.025}_{-0.016}$ at 68\% CL.
Nevertheless, considering the overall deviation of the model from $\Lambda$CDM for this dataset, the inclusion of CC with DESI+PP makes no statistical difference, as the deviation is $\sigma = 0$ (see the last column of Table~\ref{table:CPL-w0wa}).

Moving to the remaining datasets involving SDSS BAO (i.e., SDSS+SN+CC combinations), we once again find that the cosmological constant is disfavored at approximately $2\sigma$ when DESY5 or U3 is included in the combined dataset. However, in the presence of PP, this shift weakens, reducing to approximately $1\sigma$. 
The constraints on $w_0$ remain almost unchanged after the inclusion of CC, as do those on $w_a$. Similarly, the constraints on $H_0$ and $\Omega_{\rm m}$ follow the same trend as observed in the DESI+SN and DESI+SN+CC datasets, with $H_0$ values slightly shifting towards lower values while $\Omega_{\rm m}$ adjusts proportionally due to their anti-correlation.
Overall, considering the deviation from the $\Lambda$CDM paradigm as summarized in the last column of Table~\ref{table:CPL-w0wa}, the inclusion of CC does not introduce any statistically significant changes.

\subsection{Constraints including the CMB data}

Here, we present the constraints on the key parameters of the model obtained from all the combinations of CMB and low-redshift datasets. Specifically, we have explored the constraints using CMB alone, and its combination with other datasets, such as CMB+CC, CMB+BAO (DESI and SDSS), CMB+SN (DESY5, PP, and U3), and CMB+BAO+CC, CMB+BAO+SN, and CMB+BAO+SN+CC, resulting in a total of 21 cases.

\subsubsection{CMB and CMB+CC}

In the first row of Table~\ref{table:CPL-w0wa}, we present the constraints obtained from the CMB alone dataset. 
The results for CMB alone are not new, as they have already been presented in several works (see, for example, Refs.~\cite{Planck:2018vyg,Yang:2021flj,Escamilla:2023oce}). In this case, $w_0$ lies in the phantom regime at slightly more than 68\% CL ($w_0 = -1.36^{+0.24}_{-0.53}$ at 68\% CL), but it is only weakly constrained with an upper limit at 95\% CL. The parameter $w_a$, which quantifies the dynamical nature of $w(a)$, also has only an upper limit for this dataset. 

Due to the phantom nature of $w_0$, the Hubble constant $H_0$ takes a very high value with large error bars ($H_0 = 84^{+20}_{-5}~{\rm km}/{\rm s}/{\rm Mpc}$ at 68\% CL). This high value of $H_0$ leads to a lower matter density parameter of $\Omega_m = 0.208^{+0.023}_{-0.063}$ at 68\% CL, since $H_0$ and $\Omega_m$ are anti-correlated. Overall, this dataset exhibits a deviation from the standard $\Lambda$CDM model at the $1.9\sigma$ level (see also Fig.~\ref{fig:sigma}).

The inclusion of CC data significantly affects these constraints. As seen in the second row of Table~\ref{table:CPL-w0wa}, $w_0$ shifts towards $-1$, with a constraint of $w_0 = -1.03^{+0.48}_{-0.37}$ at 68\% CL for the CMB+CC combination. Meanwhile, $w_a$ is constrained at $w_a = -0.47^{+0.54}_{-1.4}$ at 68\% CL, though it remains unconstrained from below at 95\% CL. Therefore, within 68\% CL, CMB+CC allows $(w_0, w_a) = (-1, 0)$, which is statistically very close to the standard $\Lambda$CDM model with a deviation of only $0.1\sigma$ (see also Fig.~\ref{fig:2D-w0wa}).

\subsubsection{CMB+SN}

CMB in combination with any of the three SN samples (e.g., DESY5, PP, U3) influences the parameter space quite significantly. In all three cases, as shown in Table~\ref{table:CPL-w0wa}, $w_0$ is consistently found in the quintessence regime, while $w_a$ is non-null.\footnote{For the combination CMB+U3, $w_a$ is constrained at 68\% CL but remains unconstrained from below at 95\% CL.} The strength of the evidence depends on the specific dataset: $w_0$ remains in the quintessence regime at more than $2\sigma$ for CMB+DESY5 and CMB+U3, while for CMB+PP, it stays in the same regime but only slightly above $1\sigma$. 
On the other hand, $w_a$ is non-null (i.e., $w_a \neq 0$, indicating a dynamical $w(a)$) at more than $2\sigma$ for CMB+DESY5 and CMB+U3, but only at approximately $1\sigma$ for CMB+PP. These results jointly support the evidence for dynamical dark energy with a present-day quintessential nature of $w(a)$.

In the second panel from the top of Fig.~\ref{fig:whisker-w0wa}, we present the whisker plots for $w_0$ (left) and $w_a$ (right) for these datasets, offering a clear visualization of the DE equation of state and the strength of the dynamical DE evidence. Similar to the BAO+SN analysis discussed earlier, the evidence for DDE for CMB+SN is more pronounced when DESY5 and U3 are included, while it weakens significantly with PP, as also illustrated in Fig.~\ref{fig:2D-w0wa}. 

To quantify the statistical preference for the CPL model over $\Lambda$CDM, we calculate the deviation and find that CPL deviates from the standard $\Lambda$CDM model at $2.6\sigma$, $1.7\sigma$, and $0.1\sigma$ for CMB+DESY5, CMB+U3, and CMB+PP, respectively. This confirms that the maximum deviation is observed for CMB+DESY5 (see also Fig.~\ref{fig:sigma}).

Finally, focusing on $H_0$, our analysis reveals that for all three combined datasets, the estimated values of $H_0$ remain consistent with the predictions from Planck 2018 within the $\Lambda$CDM model~\cite{Planck:2018vyg}, though with slightly larger error bars. Similarly, the estimates for the matter density parameter $\Omega_{m0}$, as shown in Fig.~\ref{fig:whisker-Omegam}, are also consistent with Planck 2018 results within $\Lambda$CDM~\cite{Planck:2018vyg}.

\subsubsection{CMB+BAO}

When CMB is combined with BAO, significant changes in the constraints on the parameters are observed. We have explored two versions of the BAO data, specifically DESI and SDSS. Notably, CMB+DESI fails to constrain $w_a$ within the considered range $[-2, 2]$, providing only an upper limit, while CMB+SDSS does constrain $w_a$—a distinguishing feature between these BAO datasets, as shown in Table~\ref{table:CPL-w0wa} and the third panel (from top to bottom) of Fig.~\ref{fig:whisker-w0wa}.
Focusing on $w_0$, we observe that for CMB+DESI, it remains in the quintessence regime at slightly more than $2\sigma$, whereas for CMB+SDSS, $w_0 > -1$ is only weakly indicated at slightly more than $1\sigma$. This suggests that DESI combined with CMB enhances the evidence for a quintessential dark energy component more significantly than SDSS combined with CMB. Overall, we find a deviation from $\Lambda$CDM of $2.3\sigma$ for CMB+DESI and $0.3\sigma$ for CMB+SDSS, as also illustrated in Fig.~\ref{fig:sigma}. This indicates that DESI, when paired with CMB, shows a stronger preference for a DDE than SDSS, as can also be seen in Fig.~\ref{fig:2D-w0wa}. 

Regarding the Hubble constant, our results show consistency with the value obtained under $\Lambda$CDM, but on the lower side, while the matter density remains in agreement with all previously discussed cases, supported by larger error bars, as shown in Fig.~\ref{fig:whisker-Omegam}.

The inclusion of CC data with both CMB+DESI and CMB+SDSS datasets does not significantly alter the constraints. For example, $w_a$ continues to have only an upper limit for CMB+DESI+CC, indicating that the additional constraining power from CC is insufficient to improve the constraints in this case. Similarly, the constraints on $w_0$ and other cosmological parameters remain almost unchanged with the inclusion of CC for both datasets. Specifically, CPL remains $0.3\sigma$ away from $\Lambda$CDM for both CMB+SDSS and CMB+SDSS+CC. However, for CMB+DESI+CC, the deviation increases to $2.6\sigma$, compared to $2.3\sigma$ for CMB+DESI alone (see also Fig.~\ref{fig:sigma}).

\subsubsection{CMB+BAO+SN}

In this case, we consider six different combined analyses involving two BAO datasets (DESI and SDSS) and three SN samples (DESY5, PP, U3). We can immediately observe that the indication for DDE seen in the BAO+SN and CMB+SN combinations is further strengthened when considering the full combination CMB+BAO+SN. The 2D contour plots shown in Fig.~\ref{fig:2D-w0wa} reveal a striking deviation from the cosmological constant, consistent with findings already reported in the literature.

Focusing first on the constraints from CMB+DESI+SN datasets, we find that $w_0$ remains in the quintessence regime at more than $2\sigma$ for all three combined analyses. Notably, for CMB+DESI+U3, $w_0$ is significantly different from $-1$ ($w_0 = -0.661 \pm 0.095$ at 68\% CL) compared to the results obtained with the other two SN samples. On the other hand, $w_a$ is constrained in all cases, unlike the scenario without SN data (see Fig.~\ref{fig:whisker-w0wa} and Table~\ref{table:CPL-w0wa}). Moreover, $w_a$ is also found to be non-null at more than $2\sigma$, providing strong evidence for a dynamical evolution of $w(a)$ as described by the CPL parametrization, as further illustrated in Fig.~\ref{fig:eos-w}.

\begin{figure}
    \centering
    \includegraphics[width=1\linewidth]{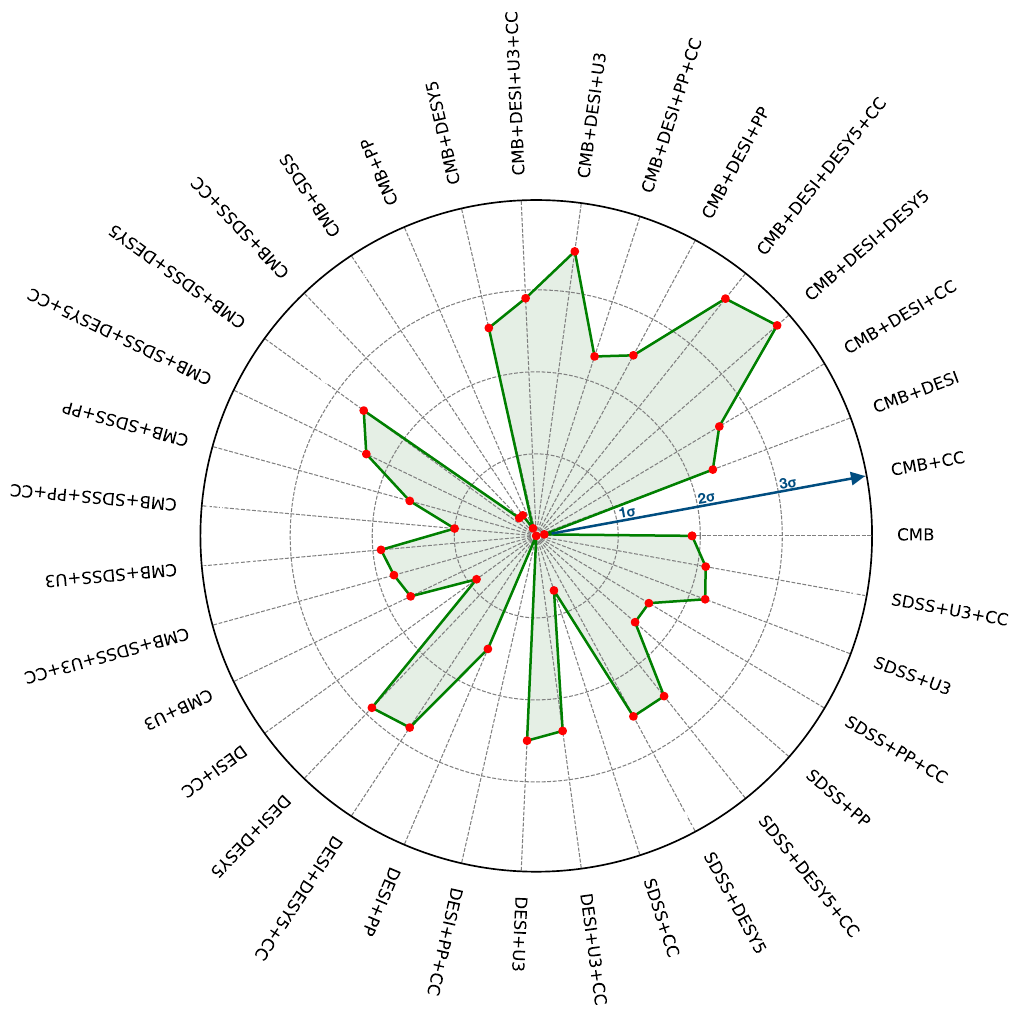}
    \caption{Radar plot quantifying the preference for the CPL evolving DE component over $\Lambda$CDM, measured by the standard deviation, $\sigma$, calculated for each observational dataset summarized in Table~\ref{table:CPL-w0wa}, following the methodology described in Sec.~\ref{sec:data+Methodology}. }
    \label{fig:sigma}
\end{figure}

When comparing with the $\Lambda$CDM model, a deviation of $3.9\sigma$ is observed for CMB+DESI+DESY5 and $3.5\sigma$ for CMB+DESI+U3, both of which are significantly larger than the $2.5\sigma$ deviation seen for CMB+DESI+PP. These cases represent the largest deviations explored in this study, as visualized in Fig.~\ref{fig:sigma}.

The Hubble constant $H_0$ is extremely well constrained in these cases and remains in agreement with the $\Lambda$CDM model, while the matter density $\Omega_m$ is slightly higher but still consistent with the standard value, as shown in Fig.~\ref{fig:whisker-Omegam}.

Upon including CC data in all three combined datasets, the constraints on $w_0$, $w_a$, and other cosmological parameters remain nearly unchanged due to the limited constraining power of CC data. Consequently, the statistical deviation of the model from $\Lambda$CDM remains almost identical to the cases without CC.

The replacement of DESI by SDSS in the CMB+BAO+SN combination significantly affects the constraints, though it still mildly favors a DDE in all cases, as shown in Fig.~\ref{fig:2D-w0wa}. For all three SN datasets, $w_a$ remains non-null at more than $2\sigma$, though the strength of evidence is more pronounced in the presence of DESI, as illustrated in the whisker plot of Fig.~\ref{fig:whisker-w0wa} and the evolution of the dark energy equation of state $w(a)$ in Fig.~\ref{fig:eos-w}. 
Regarding $w_0$, it is found to be in the quintessence regime at more than $2\sigma$ when DESY5 and U3 are considered, while for PP, the evidence is reduced to approximately $1\sigma$. A notable trend is the shift of $w_0$ towards $-1$ compared to the constraints derived from CMB+DESI+SN (see also Table~\ref{table:CPL-w0wa}). Statistically, the evidence for dynamical DE is weaker in the presence of SDSS compared to DESI. Specifically, the evidence for dynamical DE is reduced from $3.9\sigma$ (CMB+DESI+DESY5) to $2.6\sigma$ (CMB+SDSS+DESY5). Similarly, the evidence for CMB+SDSS+PP drops to $1.6\sigma$, compared to $2.5\sigma$ with DESI. For CMB+SDSS+U3, the deviation is reduced to $1.9\sigma$ from $3.5\sigma$ with DESI (see the radar plot in Fig.~\ref{fig:sigma}).

Analogously to the case with DESI, both the Hubble constant $H_0$ and the matter density parameter $\Omega_m$ are extremely well constrained for CMB+SDSS+SN and remain consistent with the $\Lambda$CDM model (see Fig.~\ref{fig:whisker-Omegam}).

When CC is included in the CMB+SDSS+SN combinations, the evidence for dynamical DE is mildly reduced across all cases. The most significant reduction occurs for CMB+SDSS+PP and CMB+SDSS+PP+CC, where the inclusion of CC lowers the evidence for dynamical DE from $1.6\sigma$ to $1\sigma$ (see also Fig.~\ref{fig:sigma}).

\section{Summary and Conclusions}
\label{sec:Conclusion} 

After the BAO measurements from the DESI collaboration~\cite{DESI:2024uvr,DESI:2024mwx}, intriguing features of the 
DE sector emerged. Assuming a dynamical DE equation of state described by the CPL  
parametrization: 
$w(a) = w_0 + w_a (1-a)$, it was reported that the combination of CMB+DESI-BAO indicates a preference for a dynamical DE component at approximately $2.6\sigma$. When various SN datasets are added to CMB+DESI-BAO, the preference for dynamical DE strengthens, reaching $3.9\sigma$ for the CMB+DESI-BAO+DESY5 combination.

This result challenges the cosmological constant interpretation of the present-day accelerated expansion of the Universe, pointing to a dynamical DE component characterized by present-day quintessence-like behavior ($w_0 > -1$) that transitioned into the phantom-like regime in the past ($w_a < 0$).

Motivated by these findings, we presented a concise review of the current constraints on dynamical dark energy. Adopting the Chevallier-Polarski-Linder parametrization to describe the evolution of the equation of state, we systematically analyzed multiple cosmological probes, considering over 35 different combinations of data involving two BAO datasets (DESI and SDSS), three SN datasets (DESY5, PP, and U3), Hubble parameter measurements from cosmic chronometers, and Planck CMB temperature and polarization anisotropies. Our goals were to provide the community with a reader-friendly synthesis of what the latest cosmological and astrophysical probes can (and cannot yet) reveal about dynamical dark energy, clarify the constraining power of each individual probe, and determine whether, and to what extent, the preference for dynamical dark energy was supported across different combinations of datasets, accounting for varying sources of information from high and low redshift surveys.

We summarized the numerical constraints on key parameters in Table~\ref{table:CPL-w0wa}, while the key results were visually presented in Figs.~\ref{fig:2D-w0wa}, \ref{fig:whisker-w0wa}, \ref{fig:eos-w}, \ref{fig:whisker-Omegam}, and \ref{fig:sigma}, using two-dimensional joint contours, several whisker plots, and a radar plot. 

The most relevant findings from this comprehensive study are outlined as follows:

\begin{itemize}

\item Our analysis of low-redshift probes revealed that an indication for dynamical dark energy is already present in the BAO+SN dataset combination. Notably, this preference is not solely driven by the DESI data, as a shift toward dynamical dark energy is also observed when considering SDSS BAO in combination with SN. However, the strength of this preference is reduced when SDSS replaces DESI in the same dataset combination. Importantly, among the BAO+SN analyses, the choice of the SN catalog plays the most significant role in determining the strength of the preference. In particular, the PantheonPlus dataset combined with SDSS BAO is the only case that does not yield a significant indication for dynamical dark energy (see also the top-left panel in Figure~\ref{fig:eos-w} and the 2D contours in Figure~\ref{fig:2D-w0wa}).  
 
\item When considering the CMB+DESI-BAO combination, the cosmological constant value falls outside the 95\% confidence level marginalized probability contours, whereas for CMB+SDSS-BAO, it remains within the probability contours (see Figure~\ref{fig:2D-w0wa}). That said, even for SDSS-BAO, the favored region of parameter space is consistent with that preferred by CMB+DESI. Projecting the constraints on $w_0$ and $w_a$ in one dimension, we still observe a shift toward a present-day quintessence-like equation of state, dynamically evolving into a phantom-like behavior in the past (see Figure~\ref{fig:eos-w}).

\item When considering different combinations of CMB and SN surveys, the preference for dynamical dark energy persists, reinforcing the crucial role of SN measurements in driving this trend. However, the inclusion of the PantheonPlus SN catalog weakens this indication compared to the DESY5 and Union3 datasets. 

\item The full combination of CMB+BAO+SN typically strengthens the evidence for DDE, reaching a maximal significance of $3.9\sigma$ for CMB+DESY5+DESI. As usual, this preference is reduced in combinations involving PantheonPlus SN data or SDSS BAO measurements, compared to the DESY5/U3 SN catalogs and DESI BAO, respectively. However, the preference for DDE remains robust across most dataset combinations. In fact, the only scenario where this preference is significantly weakened is when SDSS BAO and PantheonPlus SN are considered simultaneously.  

\item The addition of Cosmic Chronometers to combinations involving only low-redshift probes generally has a modest impact. One notable exception is the DESI+PantheonPlus combination, which favors the CPL model over $\Lambda$CDM at $\sim 1.5\sigma$. However, when Cosmic Chronometers are included, no statistically significant deviation between from a $\Lambda$CDM cosmology is found. When high-redshift CMB data are incorporated, the constraining power of the combined CMB+BAO+SN datasets is sufficient to render the addition of Cosmic Chronometers largely irrelevant in terms of constraints.

\item Across all independent correlations (each showing hints of deviation from a cosmological constant at varying statistical significance) we consistently observe shifts in parameter space in the same direction. These shifts point toward a quintessence-like equation of state in the present epoch that transitions into a phantom-like regime in the past. This trend is clearly visible in Figure~\ref{fig:2D-w0wa}, where the 2D contours in the $w_0$-$w_a$ plane illustrate the correlation, in Figure~\ref{fig:whisker-w0wa}, which presents the 1D marginalized constraints on $w_0$ (left panel) and $w_a$ (right panel), and in Figure~\ref{fig:eos-w}, which reconstructs the redshift evolution of the dark energy equation of state.  

\end{itemize}

Based on these results, one could take an optimistic perspective, arguing that the hints for DDE should be regarded positively, as they emerge from multiple independent probes that exhibit similar trends in the evolution of the dark energy equation of state—consistently favoring a phantom-like behavior in the past and a quintessence-like behavior today. However, on the other hand, the strength of this preference remains contingent on specific dataset choices, see also Figure~\ref{fig:sigma}. Further cross-checks using future surveys and new data releases will be necessary before these findings can be considered robust across all datasets.

\section*{Acknowledgments}

 W.G.\ acknowledges support from the Lancaster–Sheffield Consortium for Fundamental Physics through the Science and Technology Facilities Council (STFC) grant ST/X000621/1. E.D.V.\ acknowledges support from the Royal Society through a Royal Society Dorothy Hodgkin Research Fellowship.  S.P.\ acknowledges the financial support from the Department of Science and Technology (DST), Govt. of India under the Scheme   ``Fund for Improvement of S\&T Infrastructure (FIST)'' (File No. SR/FST/MS-I/2019/41). This article is based upon work from the COST Action CA21136 ``Addressing observational tensions in cosmology with systematics and fundamental physics'' (CosmoVerse), supported by COST (European Cooperation in Science and Technology). We acknowledge IT Services at The University of Sheffield for the provision of services for High
Performance Computing. 

\bibliography{biblio}
\end{document}